# The First Precise Determination of Graphene Functionalisation by *in situ* Raman Spectroscopy

*Philipp Vecera[1], Julio C. Chacón-Torres[2,3], Thomas Pichler[4], Stephanie Reich[2], Himadri R. Soni[5], Andreas Görling[5], Konstantin Edelthalhammer[1], Herwig Peterlik[4], Frank Hauke[1], and Andreas Hirsch[1]\**

(1) Department of Chemistry and Pharmacy and Joint Institute of Advanced Materials and Processes (ZMP), Friedrich-Alexander University of Erlangen-Nuremberg
Henkestrasse 42, 91054 Erlangen, Germany
Fax: (+) 49 (0)9131 85 26864
E-mail: andreas.hirsch@fau.de

(2) Institut für Experimental Physik, Freie Universität Berlin, Arnimallee 14, 14195 Berlin, Germany

(3) Yachay Tech University, School of Physical Sciences and Nanotechnology, 100119-Urcuquí, Ecuador

(4) Faculty of Physics, University of Vienna, Strudlhofgasse 4, A-1090 Vienna, Austria

(5) Chair of Theoretical Chemistry, Friedrich-Alexander University Erlangen-Nürnberg (FAU), Egerlandstraße 3, 91058 Erlangen, Germany

*Abstract:*

We report, for the first time, a comprehensive study involving *in situ* Raman spectroscopy supported by quantum mechanical calculations to exactly monitor the covalent binding to graphene with unprecedented precision. As model reaction we have chosen the hydrogenation of reduced graphite ($KC_8$) with $H_2O$ and compared it with the corresponding exposure to $H_2$ and $O_2$. The early stages of graphene hydrogenation are accompanied by the evolution of a series of so far undiscovered D-bands ($D_1$-$D_5$). Using quantum mechanical calculations we were able to unambiguously assign these bands to distinct lattice vibrations in the neighbourhood of the covalently bound addend. Interestingly, the exposure of $KC_8$ to $H_2$ and $O_2$ didn't cause covalent binding but intercalation of molecular $H_2$ or partial oxidation, respectively. A combination of $H_2O$ and $O_2$ treatment led to formation of additional hydroxyl (-OH) functionalities. The latter reaction represents a very suitable model for the decomposition of graphenides under ambient conditions (hydrogenation and hydroxylation). We have applied this Raman analysis to simulate and satisfactorily characterise a series of additional covalently functionalised graphene derivatives prepared as bulk materials with different composition (e.g. degree of functionalisation and nature of covalent addend) demonstrating the generality of the concept and the fundamental value for graphene chemistry.

**Introduction**

The wet-chemical exfoliation of graphite intercalation compounds (GICs) and the subsequent treatment with electrophiles is one of the most potent methods for covalent graphene functionalisation.[1-5] For this purpose graphite is typically activated by saturation doping with potassium to reach the highest stage I intercalation level with a crystalline K to C ratio of 1:8.[6, 7] In the subsequent covalent binding step, a single electron transfer to the electrophile (e.g. alkyl halide[8] or diazonium compound[4]) takes place and after halide- or $N_2$ elimination, the intermediately formed organic radical attacks the conjugated $\pi$-system of the graphenide upon the formation of $sp^3$ centres in the carbon lattice.[9-12] The degree of functionalisation (DOF) depends on the reduction potential of the electrophile and if this is low enough, almost all negative charges of the graphenide can be quenched.[13, 14] In our recent work, we were able to provide a simple and efficient procedure for the quantitative discharging of reduced graphites using benzonitrile as trapping reagent, which allows for the synthesis of defect-free graphene from graphenide solutions.[6]

Both the determination of the successful covalent functionalisation and of the quality of exfoliated graphene can be determined by Raman spectroscopy, which serves as the most important characterisation tool for the analysis of graphene based materials.[15-18] It is a non-destructive technique, allowing for disentangling the interaction between individual graphene sheets and functional groups. Raman spectroscopy and in particular statistical Raman microscopy (SRM)[9] can also be used to analyse the doping effects,[19, 20] strain,[21] oxidation and sample quality,[22-24] molecular functionalisation[25] and number of layers.[26] For this purpose characteristic changes of the most prominent Raman modes, namely, the D-, G- and 2D-modes are the most significant indicators.[15] However, a graphitic framework containing lattice embedded $sp^3$ carbon atoms – generally termed as $sp^3$ defects – gives rise to additional Raman modes.[15] Moreover, recent work predicted the presence of additional Raman bands for hydroxylated graphene,[22] which have been observed in graphene oxide (GO) samples[27]. Moreover, first approaches for the quantification of $sp^3$ defects have been reported.[28, 29] Based on these considerations we have developed a geometrical model revealing the DOF θ as ratio of the basal $sp^3/sp^2$ carbon atoms by the use of scanning Raman spectroscopy (SRS) for statistical analysis.[9, 30] However, the information obtained from Raman spectroscopy is only valid for interdefect distances ∼3 nm[28, 29] and the corresponding DOF θ < 0.5%.[9, 30] Therefore, functionalised graphene derivatives like polyhydrogenated graphane[31, 32] or graphene oxide[33, 34] still can not be addressed. In those cases, the D-, G- and 2D- modes appear as very broad and poorly resolved features[18] hiding the individual contributions from the individual lattice vibrations.[35, 36] The unequivocal assignment and resolution of individual lattice modes introduced by covalent binding is elusive and remains a major challenge in graphene chemistry. Tackling this problem would require a) the *in situ* spectroscopic monitoring of the reaction progress before the defect-induced broadening of the Raman modes in highly functionalised samples prevents any line shape analysis and b) a detailed understanding of the correlation between defect-related Raman modes and the atomic structure of the addend carrying neighbourhood in the covalent adduct. We now report for the first time a comprehensive study involving *in situ* Raman spectroscopy supported by quantum mechanical calculations where we have successfully solved the challenges pointed out above. As model reaction we have chosen the hydrogenation of reduced graphite[32] with $H_2O$ and compared it with the corresponding exposure to $H_2$ and $O_2$. Next to the very precise characterisation of the covalently functionalised graphene by an unambiguous assignment of the lattice vibrations, we were able to provide profound mechanistic information on the underlying covalent addition chemistry. Our results are of fundamental importance for any laboratory investigating the chemistry and materials design of graphene, graphene composites and other functional synthetic carbon allotropes.

**Results and Discussion**

The setup of the *in situ* Raman monitoring of the reaction of defect free $KC_8$ crystals with $H_2O$, $H_2$ and $O_2$ is presented in **Figure 1a**. This high-end system enables an unprecedented precise reaction control since a focused scenario consisting of the two reaction partners is provided. In this setup, the partial pressure of the volatile component at the solid/gas interphase is the only parameter which is varied. The *in situ* Raman setup is equipped with a laser probe (excitation wavelength 514 nm) which allows for the spectroscopic monitoring of the reaction and the detection of the stepwise evolution of functionalisation related Raman modes. The required stage I GIC ($K_8C$) was prepared under argon atmosphere.[6] The successful and clean formation of $KC_8$ was confirmed by Raman spectroscopy and by XRD analysis (Figure S 3). After controlled exposure to $H_2O$ vapour we were able to monitor the early stage of the reaction (low degree of addition) associated with a mild surface hydrogenation (**Figure 1b**). In **Figure 1c**, the evolution of the Raman spectra of a fully doped GIC ($nK^+C_8^{n-}$) is presented upon extended exposure to $H_2O$ vapour. The gradual growth of distinct modes becomes apparent.

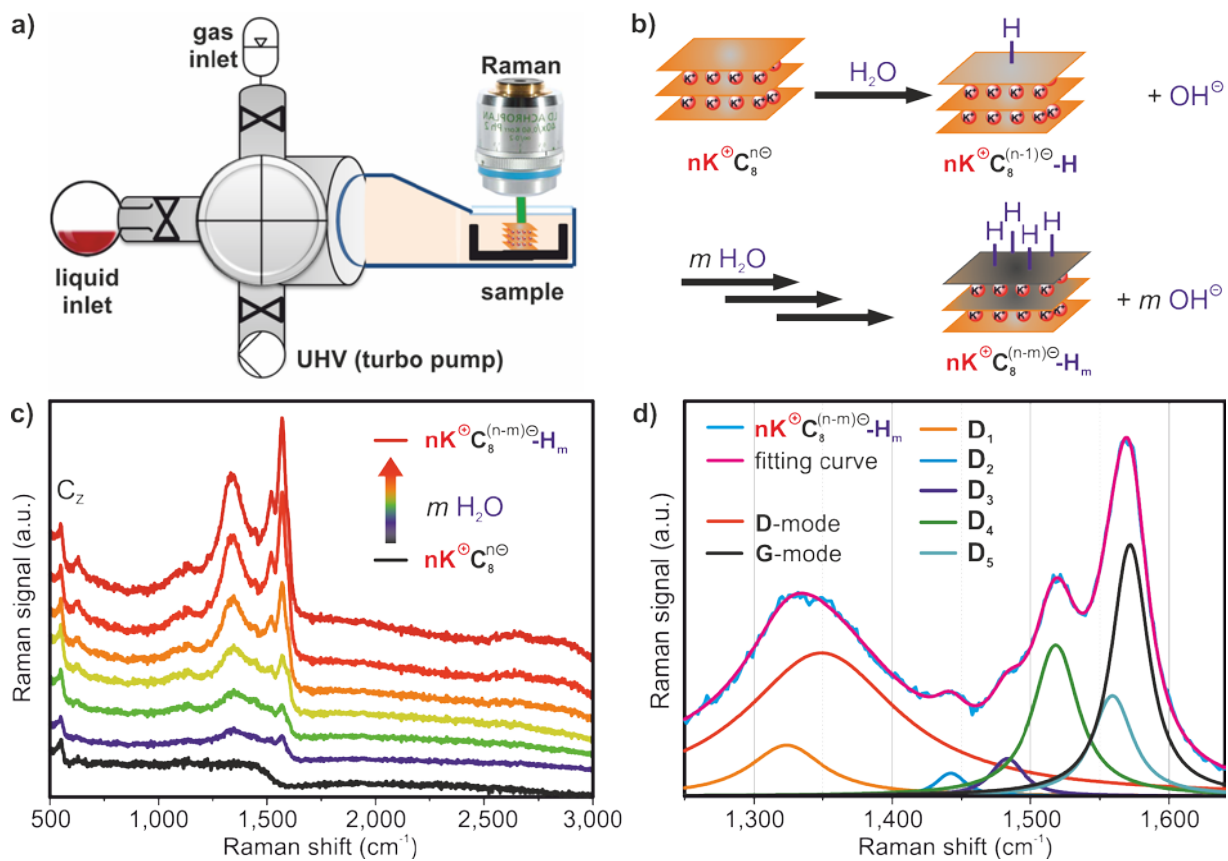

**Figure 1**: a) Schematic illustration of the setup for the controlled reaction of $KC_8$ with liquid and/or gaseous reagents under UHV conditions. The reaction progress on the surface is monitored by *in situ* by Raman spectroscopy. b) Scheme of the model reaction: Defect-free stage I GIC ($nK^+C_8^{n-}$) was exposed to $H_2O$ vapour. After the initiation of the reaction, hydrogenated graphene is formed. c) Evolution of the Raman spectra from $nK^+C_8^{n-}$ (black, bottom) to hydrogenated $nK^+C_8^{(n-m)-}H_m$ (red, top) in the first stages of $sp^3$ defect site formation on the crystal. d) Resulting Raman fingerprint of $nK^+C_8^{(n-m)-}H_m$ after $sp^3$ defect site formation within the graphene lattice. The D- and G-line region (1,200 – 1,700 cm$^{-1}$) contains 7 components (cm$^{-1}$): Graphitic $E_{2g}$ G-mode (~1,575) and defect activated D-mode (~1,340), and five additional defect modes discovered for the first time in this study - namely $D_1$ (~1,325), $D_2$ (~1,442), $D_3$ (~1,483), $D_4$ (~1,518) and $D_5$ (~1,559), which originate from the $sp^3$ defect site formation in the honeycomb lattice.

After a few minutes of reaction, the pronounced surface functionalisation is reflected by the Raman spectrum displayed in **Figure 1d**. In a detailed deconvolution of the spectra at least seven main features were revealed. The most prominent mode was assigned to the graphitic $E_{2g}$ G-mode (~1,575 cm$^{-1}$), while we attribute the slight deviation in the phonon frequency to strain present in the GIC.[21] In addition to the well-known dispersive D-mode (~1,340 cm$^{-1}$) five additional modes ($D_1$-$D_5$) were identified for the first time. These modes are assigned as $D_1$ (~1,325 cm$^{-1}$), $D_2$ (~1,442 cm$^{-1}$), $D_3$ (~1,483 cm$^{-1}$), $D_4$ (~1,518 cm$^{-1}$) and $D_5$ (~1,559 cm$^{-1}$). Their appearance is based on the change of hybridisation in the C lattice near the hydrogenated C atoms. As will be demonstrated below (**Figure 2c-d**), these modes can be assigned to specific lattice vibrations by a direct comparison with quantum mechanical calculations. The pronounced $C_z$-mode (~ 560 cm$^{-1}$) indicating graphitic intercalation architecture in the bulk crystal is widely retained (**Figure 1c**). On the other hand, the additional D-modes clearly reflect the functionalisation process on the surface. Mechanistically, single electron transfer (SET) from the GIC to water protons and subsequent addition of H-radicals to the oxidised graphene surface is assumed.[7, 21] Both the presence of a $C_z$-mode and the absence of any second order mode in the final spectrum show that GIC oxidation of the bulk crystal is not complete. Obviously, both the oxidation potential of $H_2O$ and limited mobility of $K^+$ in the inner part of the crystal are not sufficient to allow for complete bulk reoxidation but can be used for a surface or thin film functionalisation. The reaction comes to an end when a limiting stoichiometry of $nK^+C_8^{(n-m)-}H_m$ is reached, as indicated in the spectrum presented in **Figure 1d**. This *in situ* investigation allowed for a clear identification of sp$^3$ defect site formations in graphene related to Raman vibrational modes.

In another series of experiments we addressed the question how GICs respond to the exposure of $H_2$, $O_2$ and a combination of $O_2$ and $H_2O$ in order simulate the behaviour und ambient conditions. The corresponding results are shown in **Figure 2**. We expected that $KC_8$ should not give rise to covalent hydrogenation with $H_2$ gas under these conditions.[37] Indeed, as can be seen in **Figure 2b**, $H_2$ exposure does not yield any covalent binding to the graphene lattice since the Fano-shaped signature of stage I is largely preserved (**Figure 2b**, blue). The corresponding evolution of the Raman spectra (Figure S 5b) rather indicates $H_2$ intercalation, leading to $(H_2)@nK^+C_8^{n-}$. The intercalation of $H_2$ in between the graphene sheets is clearly corroborated by the increasing intensity of the $C_z$-mode. In addition, the exclusive exposure of oxygen gas to the GIC was studied (Figure S 5a). The evolution of the Raman spectra clearly underline that pure oxygen is not covalently reacting with $KC_8$ but causes a partial oxidation leading to a lower overall potassium loading. This can be clearly recognised from the final Raman spectrum in **Figure 2b** (red), where no defect site related fingerprint was observed. We assume that exposure of $KC_8$ to $O_2$ leads to the formation of potassium superoxide.[1,14] Hence, the reoxidation by oxygen without mass transport of potassium (in contrast to e.g. the oxidation in benzonitrile[6]) leads to disordered graphite, which is clearly revealed in the XRD pattern (Figure S 4).

To finally simulate ambient conditions first the same hydrogenation reaction as shown in **Figure 1** was initiated leading to partially reoxidised and covalently hydrogenated $nK^+C_8^{(n-m)-}H_m$. Then subsequent exposure to $O_2$ was carried out in the presence of water, which is required for the final oxidation with hydroxyl groups on the graphene scaffold.[6,14] The formation of –OH entities in the presence of water and oxygen has recently been reported for graphenide solutions[6] and therefore represents a major field of interest in reductive functionalisation of carbon allotropes. The *in situ* Raman analysis in **Figure 2b** clearly revealed that in this case further covalent binding is promoted. It can be assumed that after the initial treatment with $H_2O$ vapour, a water film is still absorbed on the graphitic surface. Subsequent SET processes between the GIC and $O_2$ can now be accompanied by follow up reactions with $H_2O$. In the Raman spectra this is reflected by an additional defect site related interband appearing around 1,460 cm$^{-1}$. This mode can be assigned to C-O vibrations which relate to the functionalisation with covalently sp$^3$ bound –OH groups. In agreement with literature, this Raman fingerprint has already been predicted by theory[22] and investigated for graphene oxide.[27] Importantly, this study can unambiguously verify the proposed origin of this band by theoretical calculations (**Figure 2d**) and by temperature dependent Raman spectroscopy (TDRS) (**Figure 3c**) in combination with TG-MS analysis (Figure S 8). These findings prove that the covalent hydroxylation of graphenide requires the presence of both oxygen and water,[6] which are omnipresent under ambient conditions. Hence, the sole treatment with oxygen is not sufficient, which strongly supports the reported mechanism for the hydroxyl functionalisation.[14]

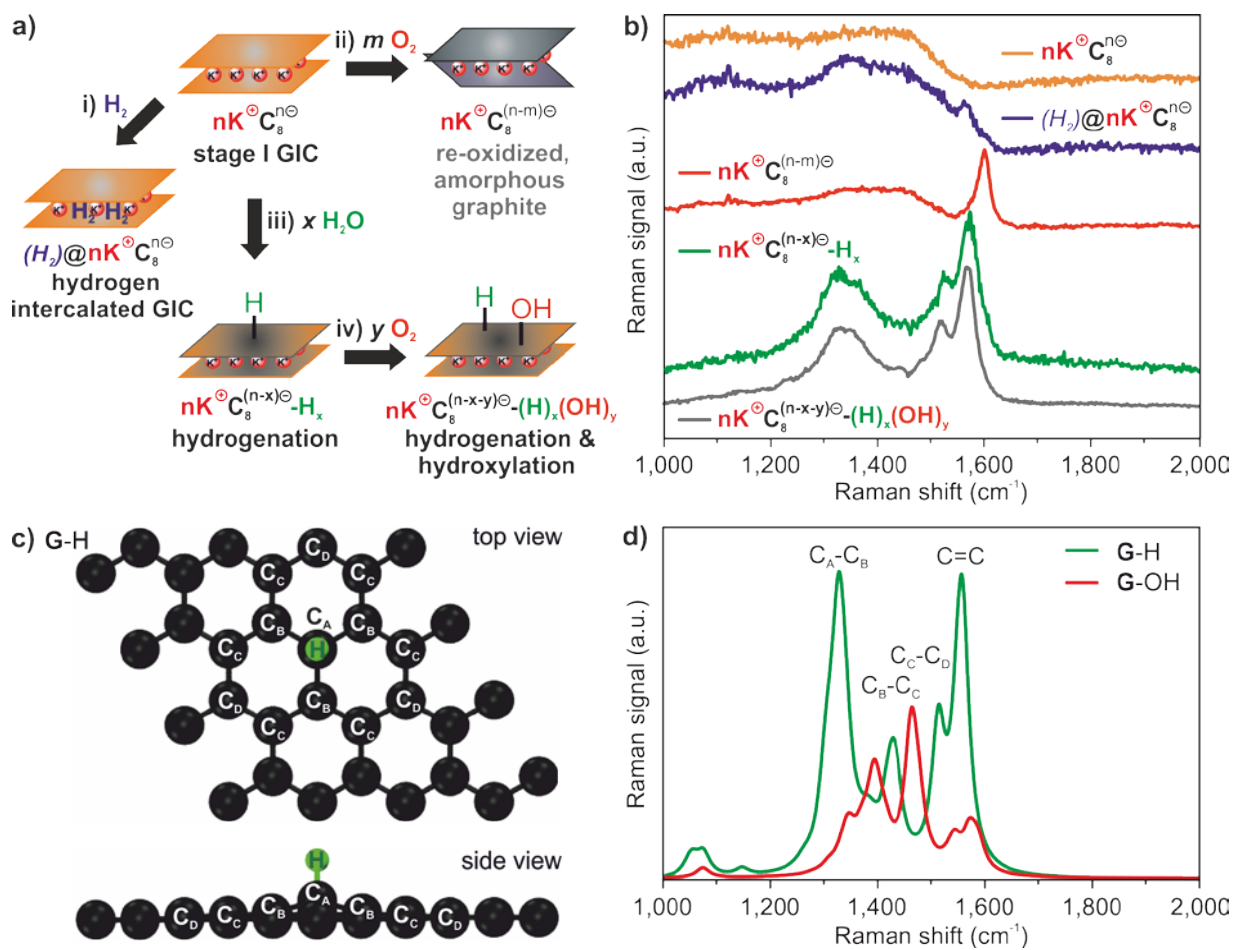

**Figure 2**: a) Systematic study of $KC_8$ exposed to i) hydrogen, ii) oxygen and iii) water which was subsequently exposed to iv) oxygen. The corresponding *in situ* Raman spectra are shown in b). Starting from saturation doped $KC_8$, i) $H_2$ leads to intercalation, ii) $O_2$ to reoxidation to amorphous graphite and iii) $H_2O$ to sp$^3$ defect site formation by hydrogenation. If iv) $O_2$ is added in combination with $H_2O$, also hydroxylation takes placed. c) 4x4 supercell of 32 lattice carbon atoms with a sp$^3$ C-H moiety attached to carbon atom A and the directly neighbouring labelled lattice carbon atoms B and C. The reference supercells for defect free graphene **G** (Figure S6) as well as for hydroxylated **G**-OH (Figure S7) are provided in the ESI. d) Calculated Raman spectra (details see supplementary information) between 1,000 - 2,000 cm$^{-1}$ with one hydrogenated / hydroxylated sp$^3$ carbon atom in a cell of 32 graphene lattice carbon atoms, respectively. The assignment of the characteristic fingerprint of sp$^3$ functionalisation is provided in Table S2.

To address the lattice carbon bond vibrational coupling deviation from the normal $E_{2g}$ G-mode related to the C=C sp$^2$ vibrations in graphene at ~1,580 cm$^{-1}$ upon introduction of sp$^3$ defect sites and the corresponding additional C-C modes, the vibrational Raman response of pristine graphene functionalised by either a hydrogen or hydroxyl addend was calculated. Therefore, a 4x4 supercell of graphene (32 lattice carbon atoms) functionalised with one -H or -OH moiety (3.125 % DOF) was considered (**Figure 2c**; computational details are provided in the ESI). The addend carrying sp$^3$ hybridised lattice carbon atom is labelled "$C_A$", while the direct neighbouring basal C atoms are termed "$C_B$" and those next to it "$C_C$" and "$C_D$". The resulting calculated Raman spectra for hydrogenated **G**-H and hydroxylated **G**-OH are presented in **Figure 2d**, respectively. The simulated reference for defect free graphene **G** is provided in Figure S 6. The calculated phonon frequencies (Tables S1 and S2) perfectly match the experimental values. In the case of **G**-H the hybridisation change is accompanied by lifting the hydrogenated C-atom by Δz = 0.46 Å out of plane (oop) in z-direction (**Figure 2c**). Along with this shift, the $C_B$-$C_A$-$C_B$ dihedral angle changes from γ = 120° (pure $sp^2$ in pristine graphene) to γ = 114.5° for **G**-H (Figure S 7a) and to γ = 113.7° for **G**-OH (Figure S 8a). The complete list of calculated angles, shifts and phonon frequencies is given in Table S1.

The theoretical analysis clearly shows that the phonon energy for excitations of lattice carbon atoms surrounding a $sp^3$ defect is affected by the newly formed C-H bond. The geometry around the C-H/$sp^3$ centre (γ ($C_BC_AC_B$) = 114.5°) is strained, since it deviates from the regular tetrahedral angle of 109.5°. The appearance of the additional $D_1$-$D_5$ modes between 1,300 and 1,600 cm$^{-1}$ is a consequence of these new geometrical constraints. When the DOF is increased, structures of curved nanodiamond clusters are eventually emerging strongly reassembling the Raman spectra of iii) and iv) in **Figure 2b**.[38] These findings are fundamental for the general interpretation of the Raman spectra of any covalent graphene derivative, since the broadening of the modes can now be precisely attributed to distinct vibrations. These results are in line with previous reports on the clustering of defect centres upon increasing DOF.[39-44] The newly observed $D_1$-$D_5$ modes start appearing in the spectra only beyond a certain DOF (θ < ~0.5 %). At even higher degrees of functionalisation (θ > 3 %), these modes broaden causing a much less structured Raman spectrum typically observed for highly functionalised graphene such as graphene oxide.[34, 45] As a consequence, the resolution and assignment of the additional modes $D_1$-$D_5$ reaches an optimum in a range of functionalisation (~0.5 % < θ < ~2 %) corresponding to the *in situ* situation depicted in **Figure 4b**.

In order to further demonstrate the importance of this powerful characterisation, we applied our Raman fingerprint assignment to crosscheck and analyse highly functionalised reaction products after workup. As a bulk functionalisation we chose the reaction of $K_8C$ dispersed in THF resulting in the exfoliation of the graphenide sheets [**G**(*THF*)]$^{ic}$ (**Figure 3a**).[2, 3] Prior to the functionalisation step, we observed no indication that $KC_8$ would undergo any reoxidation or chemisorption within the dry solvent THF. For the covalent functionalisation the intermediate [**G**(*THF*)]$^{ic}$ was subsequently exposed to oxygen and water. The resulting Raman spectrum of bulk functionalised powder sample after workup (**Figure 3b**) resembles the typical Raman signature of GO[22] where three broad overlapping modes are observed in the Raman shift regime between 1,200 – 1,650 cm$^{-1}$. In the double resonance area between 2,500 – 3,400 cm$^{-1}$ the three main components 2D, D+G and 2D* can be identified. So far these features have neither been assigned to vibrations of specific addends nor have they been used to quantify the defects in GO. We show now that the deconvolution of such spectra (**Figure 3b** top) can be accomplished and a detailed analysis of the structural composition can be provided. For this purpose the defect site related Raman fingerprint with the characteristic modes termed as D''', D'' and D' were fitted to the spectrum in **Figure 3b**. These modes are located at the same Raman shift as the previously determined interbands of charged graphite ($D_1$-$D_5$) generated *in situ* prior to workup and can therefore also be precisely identified. As indicated in **Figure 4** the $D_1$-, $D_2/D_3$- and $D_4/D_5$- modes (**Figure 1d**) can be correlated with the D''', D'' and D' interbands. Consequently, these bands can be assigned to the calculated vibrational modes $C_A$-$C_B$ ($D_1$, D'''), $C_B$-$C_C$ ($D_2/D_3$, D'') and $C_C$-$C_D$ ($D_4/D_5$, D'), respectively (**Figure 3b**). After complete reoxidation, the intravalley D*-mode could be identified, which cannot be observed in an intermediate charged state.[7, 21] The interbands $D_1$-$D_5$ of partial reacted graphene also vary in position and intensity in comparison with D''', D'' and D' of the completely reoxidised counterparts.

For an independent chemical analysis of the nature of the grafted addends, temperature dependent Raman spectroscopy (TDRS) was carried out and compared with the corresponding TG/MS results (Figure S8). This comparison allowed for the unambiguous assignment of each component in the Raman spectra. Upon thermal defunctionalisation, the defect related Raman signatures vanish at the same temperature where the –H and –OH addends are cleaved.[6]

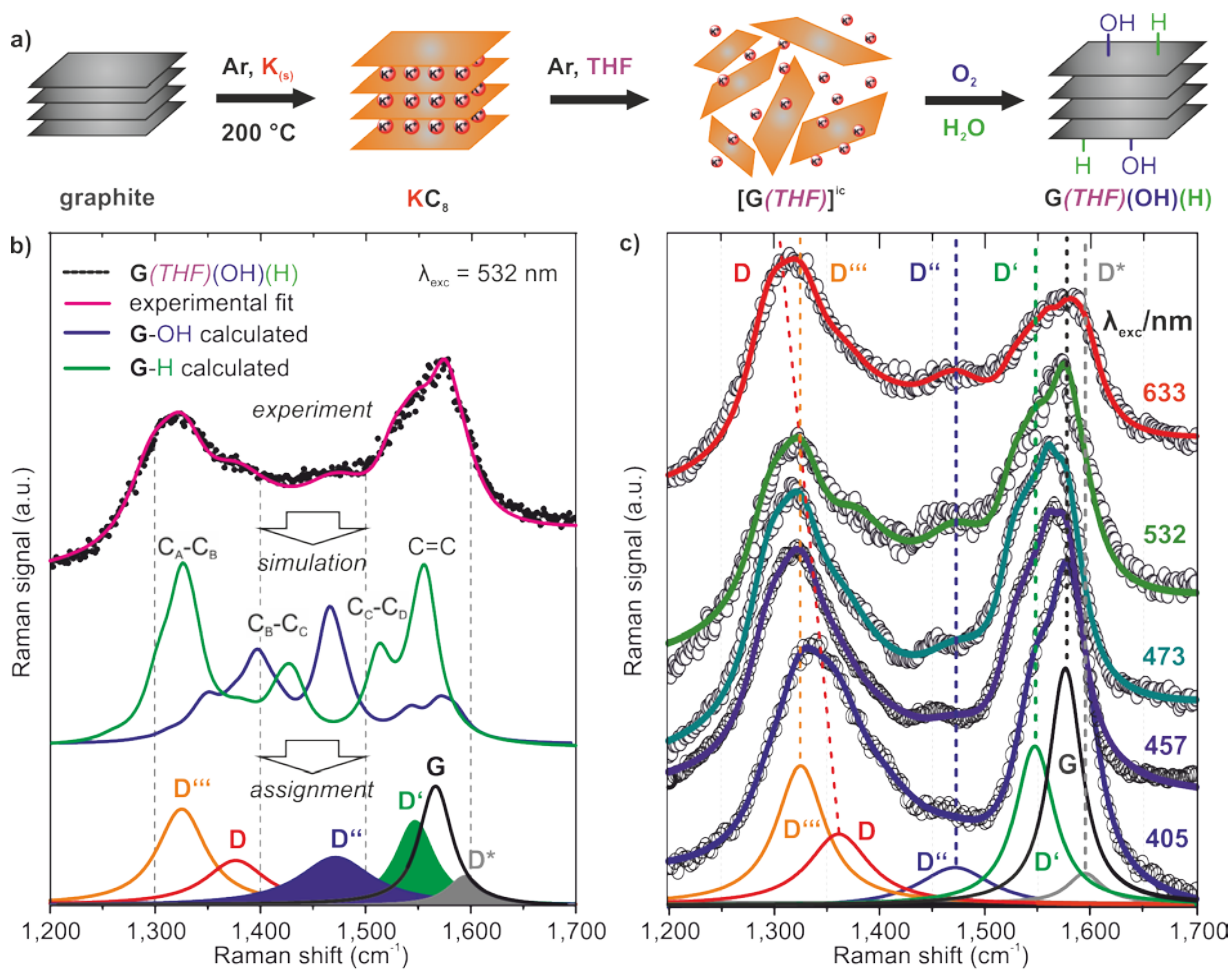

**Figure 3**: Raman spectroscopy of a bulk functionalised graphene derivative **G**(*THF*)(OH)(H) as a model of GO without σ-defects[45] a) Reaction scheme for the hydrogenation and hydroxylation of charged graphenides followed by workup under ambient conditions ($O_2/H_2O$). b) Raman D- to G-mode area in the Raman spectrum of **G**(*THF*)(OH)(H) together with the corresponding calculated spectra for **G**-OH (blue) and **G**-H (green). The combination of experiment and theory leads to a precise assignment of each Raman mode. Besides the D-, D* and G-mode arising from excitation at the k-point, the Raman fingerprint of covalently functionalised (here: -OH and –H) lattice carbon atoms can be revealed by identifying the modes D''', D'' and D'. c) Raman analysis from 405-633 nm laser excitation wavelengths, respectively. Every mode but the D-mode is non-dispersive (excitation at the Γ-point) as expected in relation to the Raman fingerprint obtained in **Figure 1**.

The development of the individual peak intensities as a function of temperature is displayed in Figure S 9b. In accordance with GO[27] (**Figure 2d**) we observe a decrease of the oxygen related D''-mode (assigned by calculations) over the whole temperature range, while the remaining functionalities – mainly hydrogen – are cleaved at higher temperatures.[31, 32] Remarkably, the evolution of this defect state related Raman mode directly correlates with the thermal cleavage of the -OH addends ($^m/_z$ 17) determined by TG-MS (Figure S 9). The dehydrogenation is clearly reflected by the $^m/_z$ 2 trace between 350 - 600 °C. The thermal dehydrogenation is accompanied by a simultaneous decrease of the D''-, D'- and the D-modes above 400 °C as depicted in Figure S 9b. Finally, to further confirm the experimental assignment of the Raman modes, a multi-frequency Raman study was conducted (**Figure 3c**). As our initial laser wavelength of 532 nm (2.33 eV) resulted in an exact superimposition of the non-dispersive D'''-mode excited at the Γ-point of the Brillouin zone (~1,325 cm$^{-1}$) and the dispersive D-mode from the k-point (~1,340 cm$^{-1}$). To prove this assumption, we decided to vary the excitation wavelength from 405 nm to 633 nm for highly functionalised samples as shown in **Figure 3c**.[15] Our results confirmed the dispersion of the D-mode in highly functionalised graphene and the non-dispersivity of the D'''-mode at ~1,325 cm$^{-1}$. To double check the experimental assignments of each component (**Figure 3b**) we carried out a cross correlation employing the calculated Raman spectra of **G**-OH and **G**-H (**Figure 3b**). Remarkably, this simulation fully matches the experimental Raman fingerprint as demonstrated in **Figure 3b**. It has to be noted that for the calculation of the Raman modes a 4x4 supercell was used. This scenario however does not reflect the Raman processes at the k-point of the graphene Brillouin zone, preventing the simulation of the D- and D*-modes. Nevertheless, all other defect site related Raman modes can be clearly assigned. The D-mode region is composed of two main components, as the non-dispersive D'''-mode and the dispersive D-mode are superimposed for λ$_{exc}$ = 532 nm. Moreover, the individual components of the Raman signal can be correlated with the TG-MS analysis (Figure S 9). Hence, this spectroscopic fingerprint is the first direct verification of the chemical nature of sp$^3$ defects (here: -OH and -H) present in the sample. The variation of the laser excitation in **Figure 3c** energy proved that none of these modes are dispersive but the Raman D-mode, which entirely agrees with our experimental and theoretical model of locally modified lattice carbon vibrations and molecular environments.

In **Figure 4**, the Raman spectra of three samples having a different DOF are presented to demonstrate the evolution of the Raman signatures with increasing $sp^3$ carbon atom content. As an example of a graphene derivative with a very low DOF ($\theta < 0.5$ %), a typical Raman spectrum of a hexyl functionalised derivative, on which we reported previously,[10] is presented in **Figure 4a**. This reductively functionalised CVD graphene sample has an isotropic distribution of defects with a distance of $L_D > 2$ nm, which results predominantly in the activation of the D- and D*-mode in the Raman spectrum. In this simple case, the $I_D/I_G$ ratio can be used for the determination of $\theta$.[9] In the case of a) the narrow width of the D-mode (22 cm$^{-1}$) relates to $\theta = 0.23$ %. At such low densities of defects, no additional Raman modes can be deduced and the deconvolution into D-, G- and D*- mode can be easily carried out. This is in very good agreement with the observed maximum in the D/G ratio in defective graphene and graphite at $L_D$ of about 3 nm[28] and 4 nm[29], respectively. The deconvolution of these modes becomes more challenging if the DOF is further increased, as the D/G ratio is reduced concomitant to a line broadening.[28,29] Our results allow for the first time to clearly resolve the origin of this broadening as additional evolving Raman interband modes. For clarity, the Raman spectrum of functionalised graphene with $\theta = 1.6$ % is shown in Figure S 10a. With ongoing functionalisation monitored by the laser probe, similar Raman interbands termed $D_1$-$D_5$ are revealed in the *in situ* reaction of KC$_8$ with H$_2$O (**Figure 4b**). The functionalisation degree of this "*in situ* resolved state" can be attributed to a range of $\theta = 0.5$ % $< \theta < 2$ %. In the corresponding covalent adducts the addends are already clustered within $sp^3$ defect site regions although more than 98 % of the basal carbon atoms are still intact. The cartoons on the bottom of **Figure 4** are presented to visualise the relationship between Raman interbands and the respective structure on the graphene sheet.

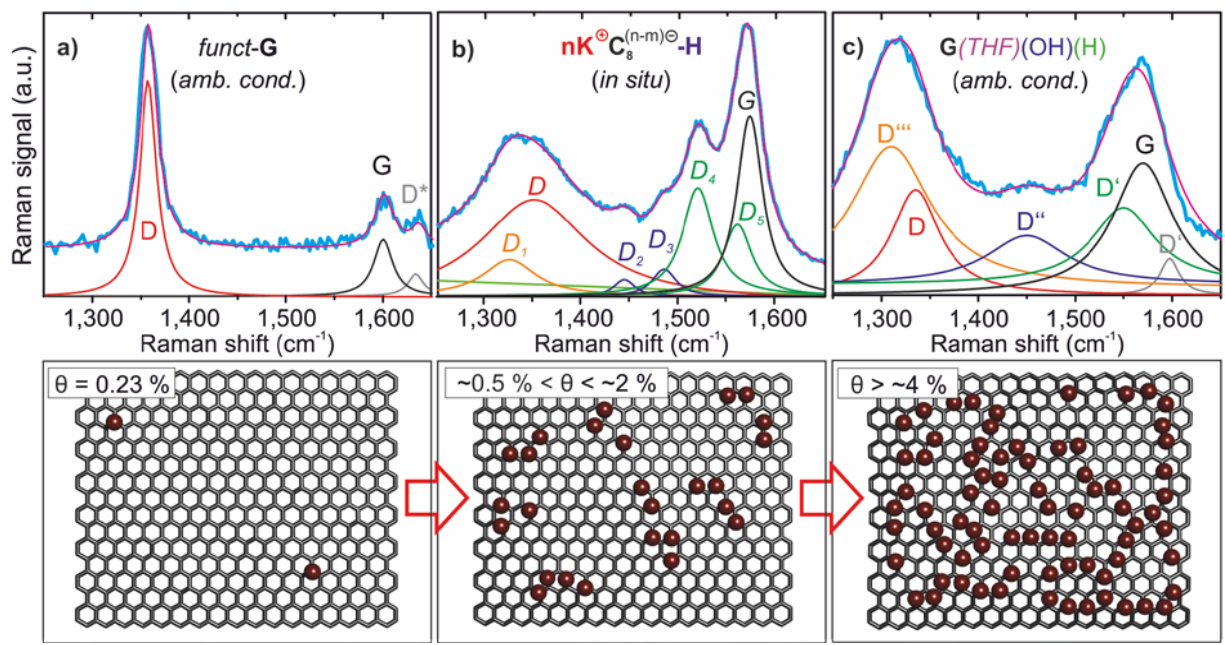

**Figure 4**: Evolution of Raman fingerprint with increasing DOF. a) Hexyl functionalised graphene[10] with isotropic distribution of functional groups with lateral DOF of $L_D > 2$ nm ($\theta < 0.5$ %). b) Local clustering of defect sites as observed in the *in situ* monitoring of the functionalisation in a typical range of $\theta = 0.5 - 2$ % using the precisely determined modes displayed **Figure 1d**. c) Higher DOF ($\theta > 2$ %) resulting in isolated $sp^2$ domains surrounded by defect-rich $sp^3$ regions.

Upon approaching the maximum DOF after full reaction of KC$_8$ under ambient conditions (**Figure 4c**), GO with $\theta = 6.0$ % serves as ideal reference,[22, 27] since the Raman modes do not change their shape but their overall intensity[30] (Figure S 10b). Thus, the analysis of all assigned Raman interbands both in GO and in **G***(THF)*(OH)(H) (**Figure 3a**) allows for the characterisation of adducts with $\theta > 2$ %. In this range, all additional components (D'''-D') in the Raman spectrum are clearly identifiable. The observed fingerprint fully matches the simulated spectra for **G**-H and **G**-OH (**Figure 3d**), which were also calculated for $\theta \approx 3$ %. This is in full agreement with recent SERS studies on functionalised CVD graphene, where a slight fingerprint for chemically modified graphene with a DOF of approx. 0.5 % leads to comparably weak Raman interbands in the D- and G-mode area.[46] Thus we can conclude that approaching the maximal possible DOF of 12.5 % (based on the ratio of K:C = 1:8) is accompanied by clustering of addends and by the formation non-altered $sp^2$ nano domains.[39-41]

**Conclusion**

We report for the first time a comprehensive study involving *in situ* Raman spectroscopy supported by quantum mechanical calculations to exactly monitor the covalent binding to graphene with unprecedented precision. This approach is very general and also allows for the fast screening and evaluation of suitable reaction conditions for covalent graphene functionalisation. As model reaction we have chosen the hydrogenation of reduced graphite ($KC_8$) with $H_2O$ and compared it with the corresponding exposure to $H_2$ and $O_2$. The early stages of graphene hydrogenation are accompanied by the evolution of a series of so far undiscovered D-modes ($D_1$-$D_5$). Using quantum mechanical calculations we were able to unambiguously assign these bands to distinct lattice vibrations in the neighborhood of the covalently bound addend. Interestingly, the exposure of $KC_8$ to $H_2$ and $O_2$ didn't cause covalent binding but intercalation of molecular $H_2$ or partial oxidation, respectively. A combination of $H_2O$ and $O_2$ treatment led to the formation of additional hydroxyl (-OH) functionalities which were clearly identified by Raman spectroscopy and TGA/MS. The latter reaction represents a very suitable model for the decomposition of graphenides under ambient conditions (hydrogenation and hydroxylation). This important process has so far never been analysed in detail. We have further demonstrated that our fundamental mechanistic investigation brings us into the position to simulate and assign the spectroscopic signatures of both, bulk functionalised **G***(THF)*(OH)(H) and graphene oxide (GO).[34] Finally, we have also applied our concept to simulate and characterise additional covalently functionalised graphene derivatives prepared as bulk materials with different composition (e.g. DOF and nature of covalent addend) demonstrating the generality of the method. So far, covalent graphene functionalisation remained a very difficult field of synthetic chemistry. This is not only because suitable methods enabling graphite/graphene activation and dispersion had to be identified in order to allow efficient adduct formation. A major challenge was also the satisfactory characterisation of reaction products since the typical powerful tools applied in synthetic chemistry such as NMR spectroscopy and mass spectrometry cannot be applied to this polydisperse 2D material. In this regard, the work presented here is a major breakthrough as it allows for graphene-derivative characterisation with unprecedented precision.

**Methods (max. 800 words)**

**Raman Spectroscopy.** *In situ* Raman spectroscopic detection (**Figure 1**) was carried out inside a quartz tube through a flat (0.7 mm thick) optical window of borosilicate glass (PGO GmbH) in a high vacuum (UHV) chamber at ~4x10$^{-8}$ mbar where the intercalated GIC was placed in a sample boat. The Raman measurements were performed at room temperature using a HORIBA LabRam spectrometer with a 514 nm excitation wavelength at 0.5 mW between 300 and 3,000 cm$^{-1}$. To avoid laser induced deintercalation and photochemistry, the laser power was kept below 0.5 mW.

The final Raman spectroscopic characterisation of the sample exposed to ambient conditions and workup (**Figure 3**) was carried out on a Horiba Jobin Yvon LabRAM evolution confocal Raman microscope (excitation wavelengths: 405, 457, 473, 532 and 633 nm) with a laser spot size of ~1 μm (Olympus LMPlanFl 50x, NA 0.50). Raman measurements were carried out using a micro-Raman setup in backscattering geometry. A charge coupled device is used to detect the signal after analysing the signal *via* a monochromator. The spectrometer was calibrated in frequency using a HOPG crystal.

**Glovebox.** Sample preparation, solvent processing and bulk functionalisation was carried out in an argon filled Labmaster SP glovebox (MBraun), equipped with a gas filter to remove solvents and an argon cooling systems, with an oxygen content <0.1 ppm and a water content <0.1 ppm.

**XRD.** X-ray diffraction was performed by placing the material in a glove box into glass capillaries with 1.5 to 2 mm diameter and 10 micron wall thickness (Hilgenberg, Germany) and subsequent sealing. X-ray patterns were measured using a microfocus X-ray source with a copper target (λ = 1.542 Å) equipped with a pinhole camera (Nanostar, Bruker AXS) and an image plate system (Fujifilm FLA 7000). All two-dimensional WAXS patterns were radially averaged and background corrected to obtain the scattering intensities in dependence on the scattering angle 2θ.

**Chemicals and solvents:**

**Graphite.** As starting material a spherical graphite SGN18 (Future Carbon, Germany), a synthetic graphite (99.99 % C, <0.01 % ash) with a comparatively small mean grain size of 18 µm (Figure S1), a high specific surface area of 6.2 $m^2$/g and a resistivity of 0.001 Ωcm was chosen. An average Raman $I_D/I_G$ intensity ratio of 0.3 is present in the starting material (Figure S2).

**Potassium chunks** were bought from Sigma-Aldrich Co. and used as received (99.99 % purity).

**Oxygen 4.5 ($O_2$) and Hydrogen 5.0 ($H_2$)** were received as lecture gas bottles (Minican) from Linde and directly connected to the in situ Raman measurement setup.

**Water ($H_2O$)** was received from Sigma-Aldrich purified, deionised and bidistilled. Pump-freeze technique was carried out 3 times to completely remove gases from the water.

**Tetrahydrofuran (THF)** was received anhydrous from Sigma-Aldrich Co. and dried over molecular sieves (3 Å). Subsequently, it was distilled over Na/K alloy to remove inhibitor and achieve absolute quality (<1 ppm $H_2O$, <1 ppm $O_2$). Finally, pump-freeze technique was used to completely degas the solvents prior to the reaction.

**Synthesis:** For the synthesis of solid state GIC mother batch, 480 mg (40 mmol carbon) spherical graphite SGN18 and 195 mg (5 mmol) potassium were heated to 200 °C in a glass vial in the glovebox. The formation of the final stage I intercalation compound was verified by *in situ* Raman spectroscopy (**Figure 1c**) and XRD analysis (Figure S 3c) under inert conditions, respectively. After the complete formation of the stage I K GIC, the powder was allowed to cool to ambient temperature and evacuated to UHV conditions prior to the reoxidation experiments carried out in the *in situ* spectroscopy setup.

The vapour pressure controlled exposure of $H_2O$, $O_2$ and $H_2$ was carried out in the specialised setup in **Figure 1a**. To achieve an efficient monitoring of the reaction between $KC_8$ and the respective reagent, the reservoir valve was opened until the pressure in the chamber was raised from $10^{-8}$ to $10^{-5}$ mbar. To further increase the concentration of reagent we stepwise allowed an increase to normal pressure for a complete floating of the sample by the reagent.

Workup of the samples at ambient conditions (**Figure 3**) for the synthesis of **G***(THF)*(OH)(H): For the Raman analysis after workup, 5 mg of $KC_8$ were dispersed in purified THF and subsequently exposed to oxygen and water under ambient conditions. After 1 h of reaction time, the black powder was washed with 10 mL of cyclohexane, ethanol and water to remove salts and solvent residues, respectively. For the final Raman analysis, the sample was dried at 70 °C overnight.

The functionalised graphene derivatives *funct*-**G** (**Figure 4**, Figure S 10a) as well as graphene oxide (GO) were produced, characterised and fitted according to literature.[30]

**Corresponding Author**

*\* E-Mail: Andreas.Hirsch@fau.de Department of Chemistry and Pharmacy, Henkestrasse 42, D-91054 Erlangen, Germany*

**Acknowledgements**

The authors thank the Deutsche Forschungsgemeinschaft (DFG-SFB 953 "Synthetic Carbon Allotropes", Project A1, Project C2), and the Graduate School Molecular Science (GSMS) for financial support. The research leading to these results has received partial fund-ing from the European Union Seventh Framework Programme under grant agreement no.604391 Graphene Flagship. J.C.T. and S.R. acknowledge the financial support of the DRS Postdoc Fellowship Point-2014 of the NanoScale Focus Area at Freie Universität Berlin. T.P. thanks the H2020 FET-Open project "2D-ink" for support.

**Author Contributions**

F.H., T.P., S.R. and A.H. supervised the project as scientific group leader and principal investigator. T.P., J.C.T. and P.V. developed the Raman setup for stable measurement conditions. T.P. provided helpful input for the treatment and interpretation of GICs. P.V. developed the concept, synthesised the GICs, performed the experimental work, processed the data and created all graphs in the manuscript. J.C.T. conducted the Raman experiments, fitted all graphs and supported the Raman characterisation. H.S. and A.G. performed the *ab initio* calculation with the assignment of bond vibrations to the Raman spectrum. K.E. purified all chemicals and supported the measurements. H.P. performed the XRD experiment and analysis. P.V., J.C.T. and A.H. wrote the manuscript.

**Supplementary Information**

Supporting information and chemical compound information are available at www.nature.com/naturechemistry. Reprints and permission information is available online at http://npg.nature.com/reprintsandpermissions/. Correspondence and requests for materials should be addressed to Andreas Hirsch.

# Supplementary Information: The First Precise Determination of Graphene Functionalisation by *in situ* Raman Spectroscopy

**Supplementary methods:**

**Computational Details.** Density-functional calculations were carried out with the Vienna *ab initio* Simulation Package (VASP)[1], which employs a plane-wave basis set. We have used "hard" pseudopotentials with a smaller core region to allow for more flexibility in the description of the valence electrons. The exchange-correlation functional due to Perdew-Burke-Ernzerhof was employed.[2] An energy cutoff of 600 eV was used. Electronic structures and geometries were converged below $1\times10^{-8}$ eV and 0.001 eV/Å with respect to total energies and forces acting on ions, respectively. We have applied a slab approach with vacuum layers of 15 Å in order to decouple periodic images from each other along the *z* direction. The Brillouin zone (BZ) was sampled by 5×5×1 Monkhorst-Pack **k**-point grids[3] for hexagonal (4×4) unit cells with 32 carbon atoms.

Vibrational frequency calculations were performed using the finite difference method. Raman intensities are calculated from the change in polarizability upon following the Eigen mode of the phonon.[4] This is calculated using the finite difference method with backward and forward calculations of each Eigen mode displacement. The dielectric tensor is reduced to a scalar in the far-from-response Raman approximation. The gamma point centred phonon modes weighted by the computed spectral intensity, convolved with a Gaussian function with a full width at half maxima of 5 cm$^{-1}$ applied to all frequencies (for comparison with experimental results), is shown in Figures S6, S7 and S8. Visualization of the calculated frequencies was performed using QVibeplot.[5]

**Supplementary Figures:**

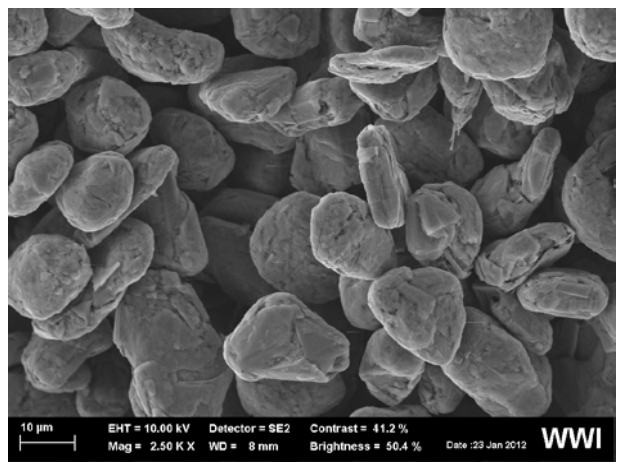

**Figure S 1**: SEM image of pristine material SGN18 graphite under 2,500x magnification.

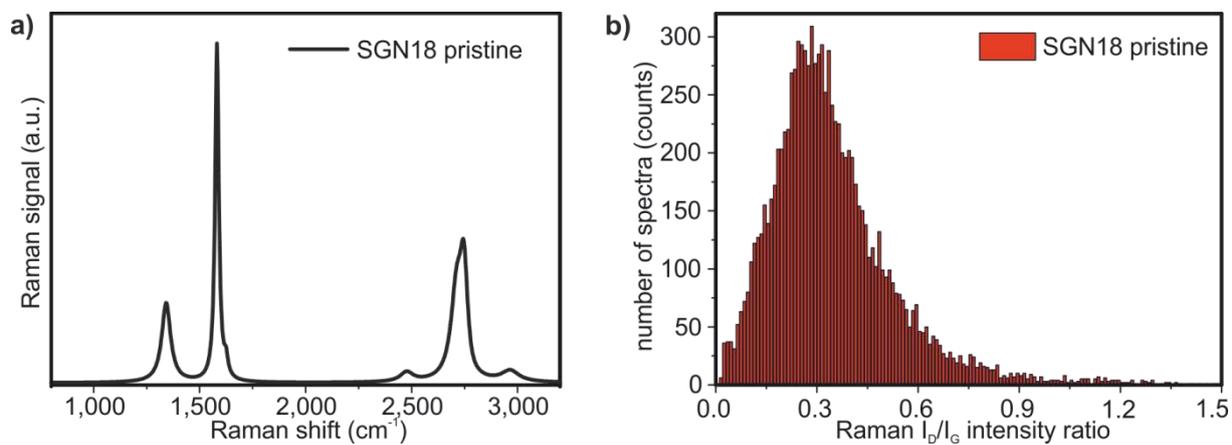

**Figure S 2**: SRS analysis of the pristine graphite material SGN18 serving as reference. 10,000 spectra were recorded leading to the mean Raman signature (left) and an $I_D/I_G$ intensity distribution given in the histogram right.

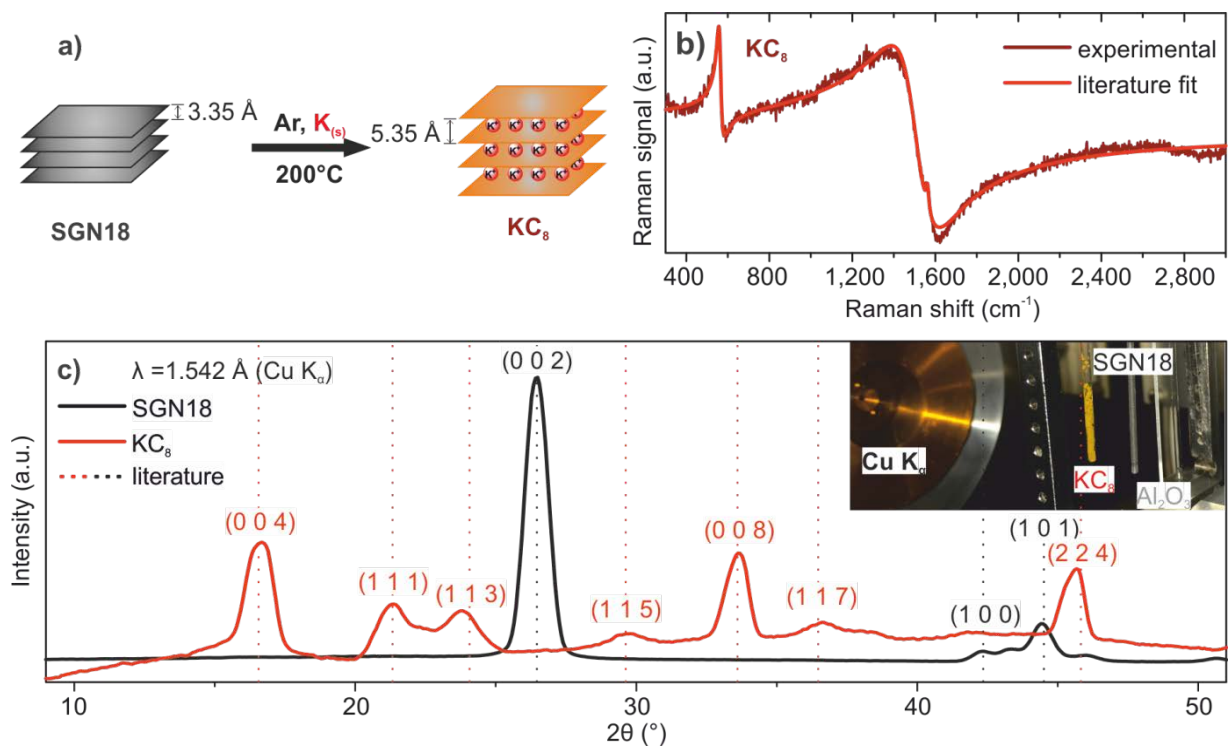

**Figure S 3**: a) Synthesis of the stage I GIC by potassium intercalation into the SGN18 graphite. b) In situ Raman spectrum of $KC_8$ indicating saturation doping. c) XRD pattern of pristine graphite (black) and $KC_8$. The interlayer distance between the sheets was determined to rise from graphite 3.35 Å to 5.35 Å for $KC_8$. The black dashed lines are literature values for the position of reflections of graphite,[6] the orange dashed lines for $KC_8$ (full orthorhombic model[7]).

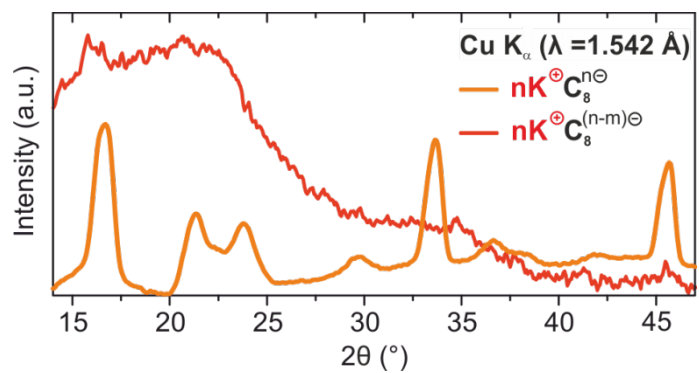

**Figure S 4**: XRD pattern of stage I GIC $KC_8$ under inert conditions (orange) and after exposure to oxygen (c.f. **Figure 2a-b**). The reflections after reoxidation reveal no crystallinity in the amorphous pattern. The X-ray pattern predominantly shows amorphous carbon, with some tiny reflections indicating relicts of intercalated potassium with approximately 5.6 Å lattice distance. It has to be noted that a small unreacted fraction of $KC_8$ in the bulk sample can be revealed by the small reflections on top of the broad reflection in the short range of 2θ.

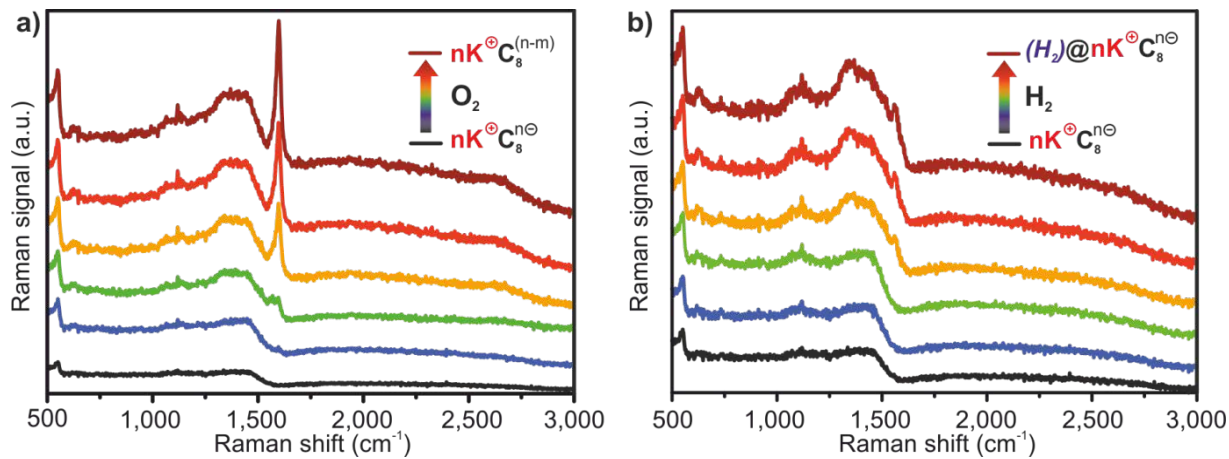

**Figure S 5**: *In situ* Raman spectroscopy of KC$_8$ exclusively exposed to a) molecular oxygen O$_2$ and b) molecular hydrogen H$_2$. The Raman spectra indicate a) oxidation of the GIC for oxygen and b) intercalation of molecular hydrogen without covalent addend binding, respectively.

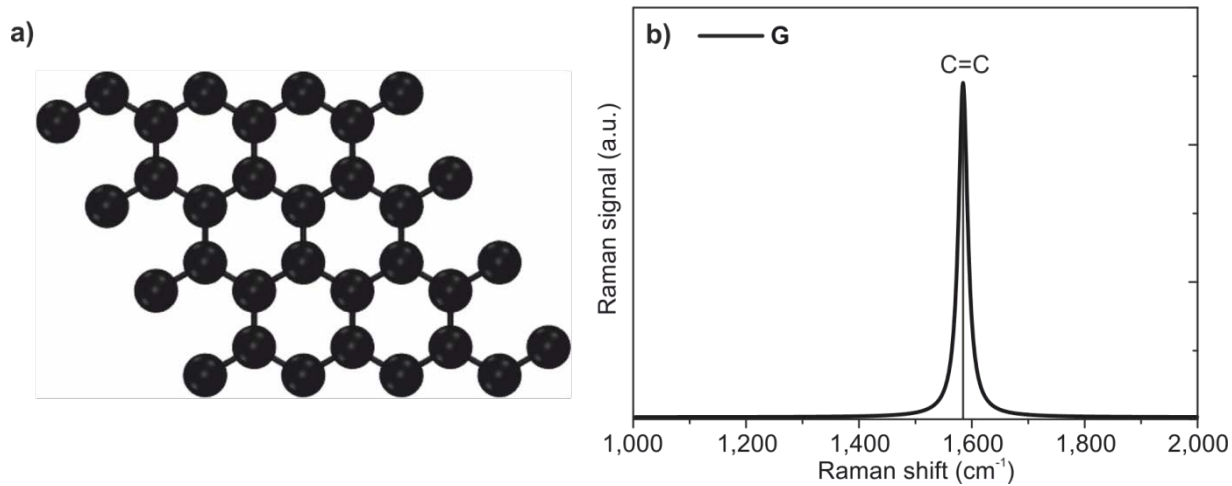

**Figure S 6**: a) Reference pristine graphene (**G**) unit cell. b) Respective calculated Raman spectrum for pristine graphene **G** without defects.

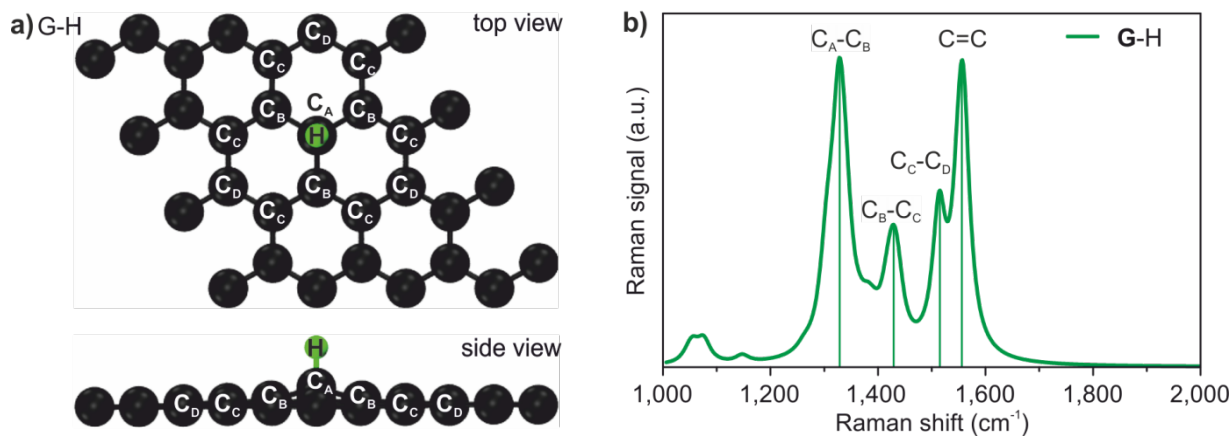

**Figure S 7:** Model for the simulation of hydrogenated graphene (**G**-H) obtained similar to **G**-OH and reference **G**. The picture shows the geometrical optimized structure from a) top view indicating the basal C atom assignment surrounding the sp$^3$ carbon atom functionalized by –H. b) Side view indicating the shift in z-direction due to the sp$^3$ hybridization. c) Calculated Raman spectrum indicating the typical fingerprint with respect to the lattice carbon atoms influenced by the sp$^3$ functionalization with –H. For the detailed vibrational frequencies and a visualization of the modes see Tables S1 and S2.

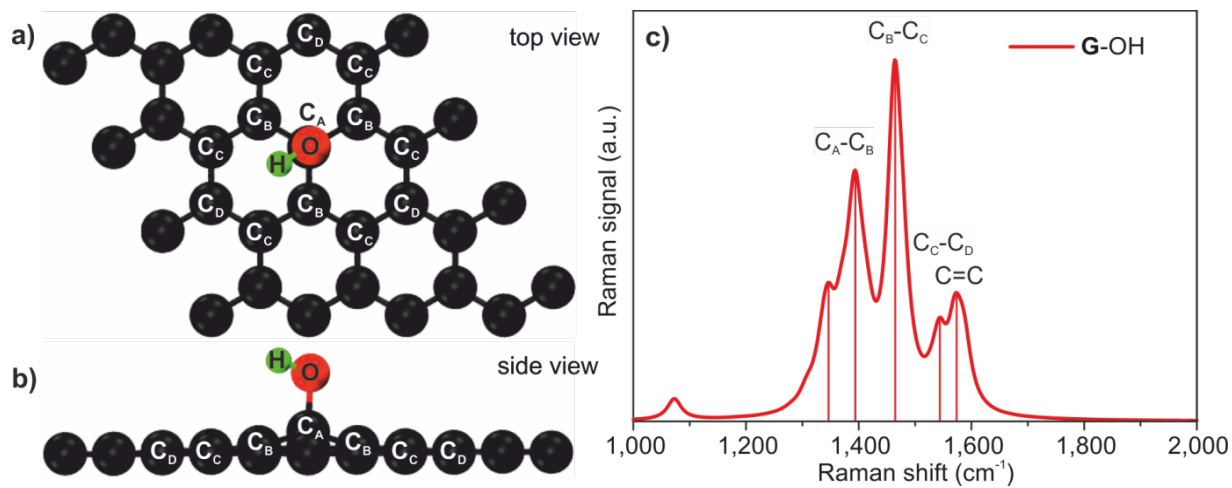

**Figure S 8**: Unit cell for the simulation of hydroxylated graphene (**G**-OH) analogous to the one used for **G**-H and reference pristine **G**. The picture shows the optimized geometric structure. a) top view indicating the basal C-atom assignment surrounding the sp$^3$ carbon atom functionalized by –OH. b) Side view indicating the shift in z-direction due to the sp$^3$ hybridization. c) Calculated Raman spectrum indicating the typical fingerprint originating from the lattice carbon atoms influenced by the sp$^3$ functionalization with –OH. For the detailed vibrational frequencies and a visualization of the modes see Tables S1 and S2.

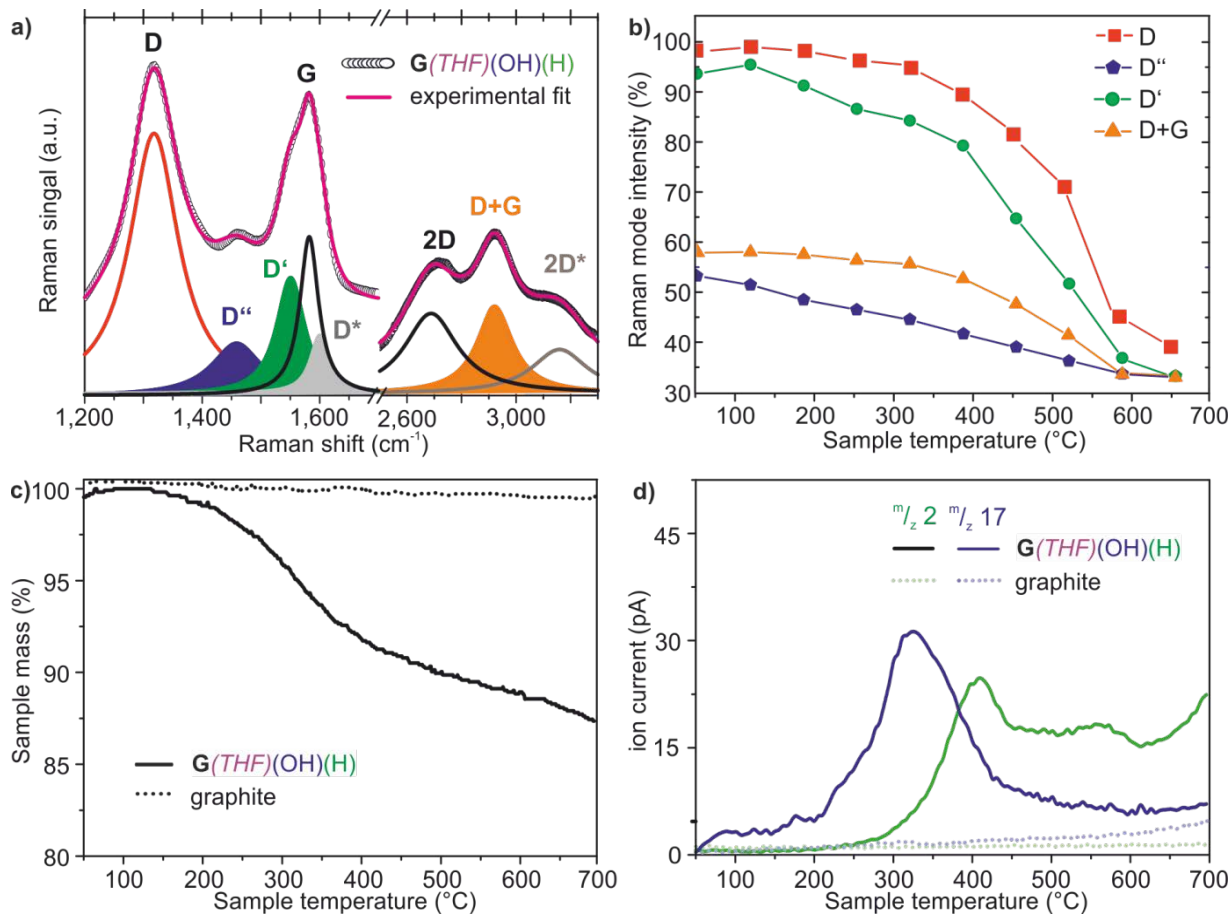

**Figure S 9**: TG-MS and TDRS analysis of the sample in Fig. 3. a) Raman spectrum of **G**(*THF*)(OH)(H) with respective fitted modes. b) Evolution of main Raman modes upon TDRS analysis. c) Weight loss curve of pristine SGN18 (-0.1 %) and **G**(*THF*)(OH)(H) (-15 %) by annealing under inert gas atmosphere (He) to 700 °C. d) Correlating MS traces for $^m/_z$ 2 ($H_2$) and $^m/_z$ 17 (OH). The TG-MS coupling proves the detachment of hydroxyl (200-400 °C) and hydrogen (350-600 °C) moieties from the sample upon heating.

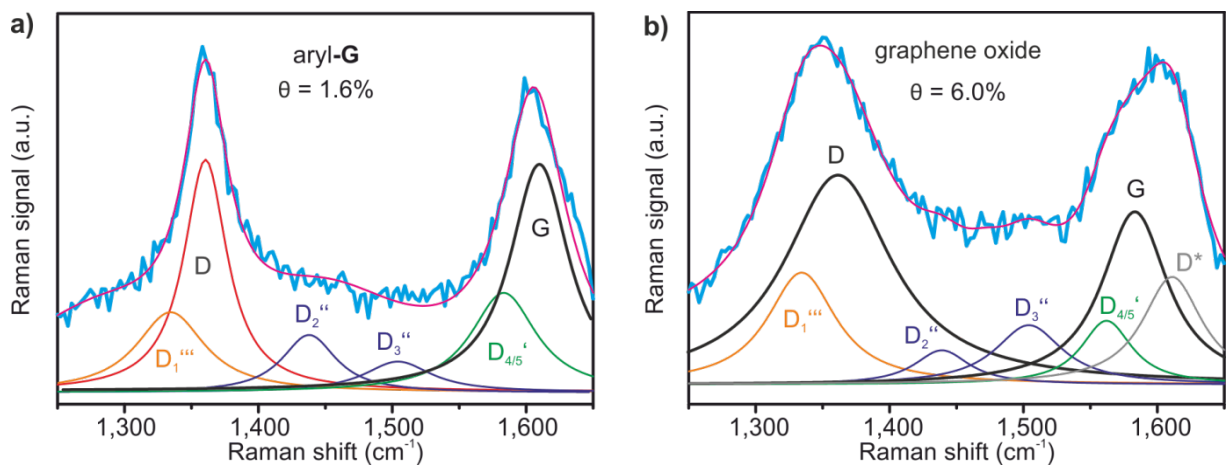

**Figure S 10**: Fitted Raman spectra for single layers of a) arylated graphene[5] and b) graphene oxide.[5] The fitting was carried out similar to **Figure 4** in the manuscript. The identified components of the spectrum perfectly match the obtained Raman modes and can therefore be correlated. Note that for aryl-G and GO, the second D'' mode can be identified.

**Supplementary Tables:**

**Table S1**: Calculated data for pristine graphene **G**, hydrogenated **G**-H and hydroxylated **G**-OH functionalized derivatives (methods see ESI). The values indicate the change in dihedral angle γ, the shift of the functionalized sp$^3$ carbon atom out of the basal plane Δ z as well as the calculated phonon frequencies for the lattice carbon atoms influenced by the introduced addend.

| graphene derivative | γ ($C_BC_AC_B$) | Δ z / Å | C=C / cm$^{-1}$ | $C_C$-$C_D$ / cm$^{-1}$ | $C_B$-$C_C$ / cm$^{-1}$ | $C_A$-$C_B$ / cm$^{-1}$ | C-X / cm$^{-1}$ |
|---|---|---|---|---|---|---|---|
| **G** | 120° | / | 1572 | / | / | / | / |
| **G**-H | 114.5° | 0.46 | 1574 | 1427 | 1348 | 1324 | 2641 |
| **G**-OH | 113.7° | 0.60 | 1576 | 1464 | 1393 | 1346 | 3614 |

**Table S2**: Simplified visualization of calculated vibrational modes using QVibplot.[5] The thickness of the lines representing the individual modes is proportional to the stretching amplitude of the respective mode. The phase of the displacement is reflected by different colors.

| Graphene derivative | **G**-H | **G**-OH | Description of the mode |
|---|---|---|---|
| C=C | | | symmetric framework stretching |
| $C_C$-$C_D$ | | | symmetric framework stretching |
| $C_B$-$C_C$ | | | asymmetric / symmetric C2 mode |
| $C_A$-$C_B$ | | | asymmetric stretching mode |

| C-X | 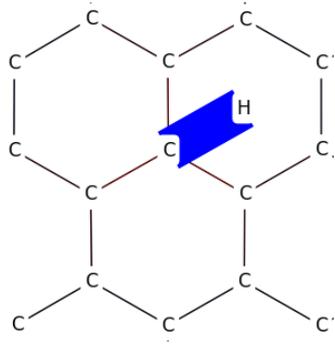 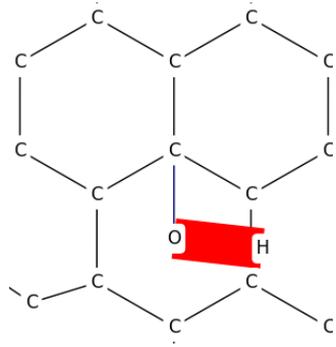 | C-H / C-OH stretching vibrations |

**Supplementary References:**